%% file: main.tex
\documentclass{vgtc}                          

\graphicspath{{figures/}{pictures/}{images/}{./}} 

\usepackage{times}                     

\usepackage{tabu}                      
\usepackage{booktabs}                  
\usepackage{lipsum}                    
\usepackage{mwe}                       

\usepackage{mathptmx}                  
\usepackage{color} 

\newcommand{\eg}{{e.g.,}\xspace}
\newcommand{\cf}{{c.f.}\xspace}

\definecolor{rtgreen}{RGB}{2,177,127}
\newcommand{\field}[1]{\textcolor{rtgreen}{#1}}

\newcommand{\octomoji}{\textsc{Emoji Encoder}}
\onlineid{0}

\vgtccategory{Research}

\vgtcinsertpkg

\title{The Data-Wink \raisebox{-.125em}{\includegraphics[height=1em]{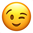}} Ratio: \\ \octomoji~for Generating Semantically-Resonant Unit Charts}

\author{Matthew Brehmer\thanks{e-mail: mbrehmer@uwaterloo.ca}\\ %
        \scriptsize University of Waterloo %
\and Vidya Setlur\thanks{e-mail: vsetlur@tableau.com}\\ %
     \scriptsize Tableau Research %
\and Zoe\thanks{e-mail: zoezoezoe.cc@gmail.com}\\ %
     \scriptsize McGraw Hill %
\and Michael Correll\thanks{e-mail: m.correll@northeastern.edu}\\ %
     \parbox{1.4in}{\scriptsize \centering Northeastern University}}

\teaser{
  \centering
  \vspace{-0.5em}
  \includegraphics[width=\linewidth]{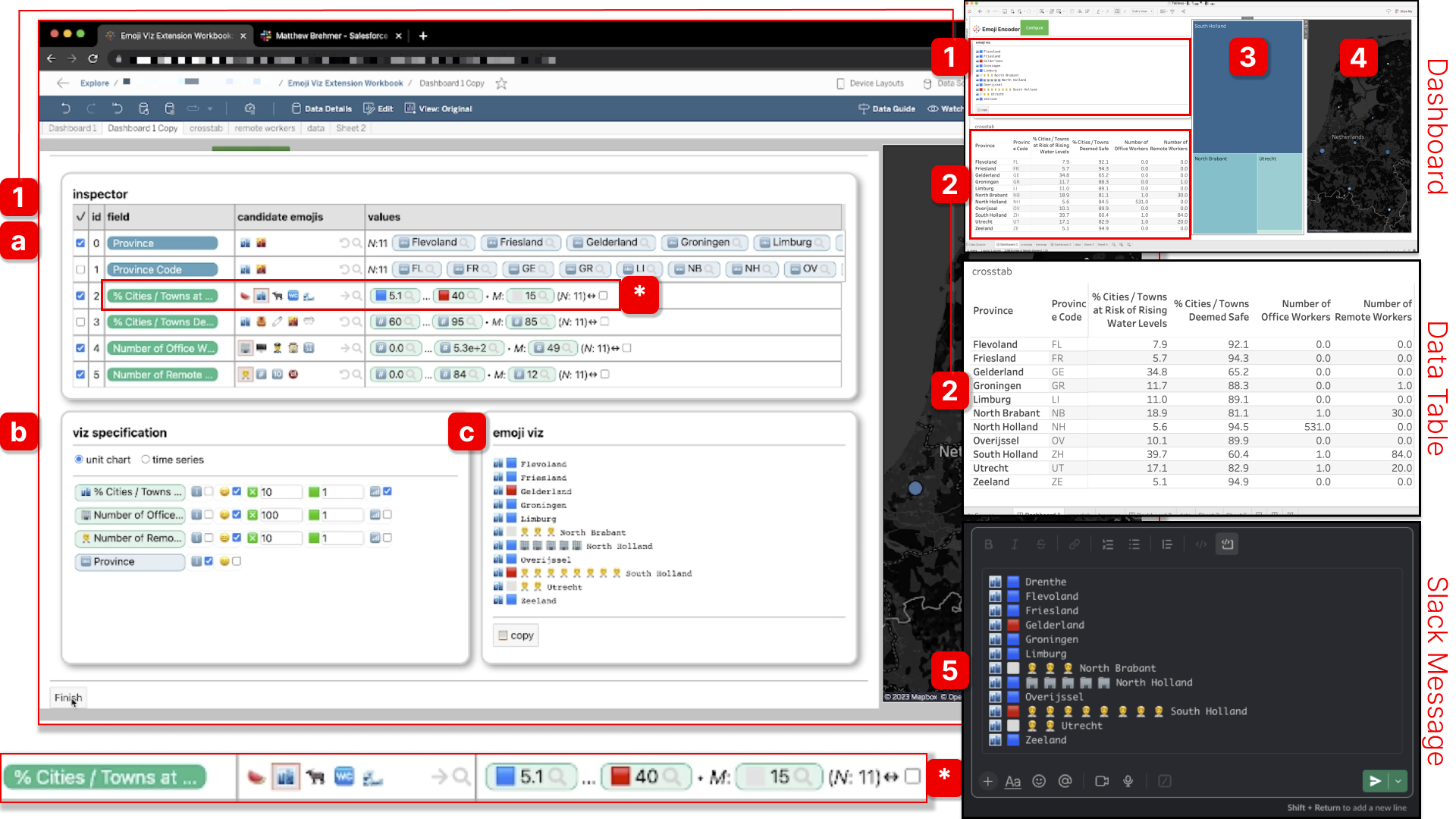}
  \caption{Beginning with the Dashboard in the top right (containing components 1 \-- 4), the \octomoji~is an interactive chart authoring interface (1) for Tableau that generates emoji representations based on field names and the values of categorical fields. In this example, a Data Table (2) contains flood risk values across the Netherlands along with the number and type of employees in each province. In the expanded inspector panel (1a), the author evaluates and selects from emoji recommendations for fields and values. For instance (*), \raisebox{-.125em}{\includegraphics[height=1em]{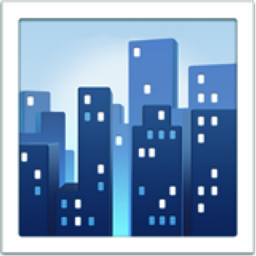}} and \raisebox{-.125em}{\includegraphics[height=1em]{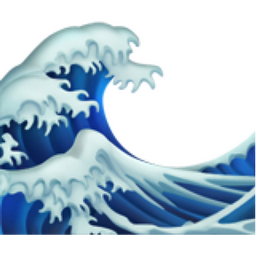}} are among the recommendations for the \field{\texttt{\% cities / towns at risk of rising water levels}} field, for which the author has selected a diverging emoji scale for its values: 
  \raisebox{-.125em}{\includegraphics[height=1em]{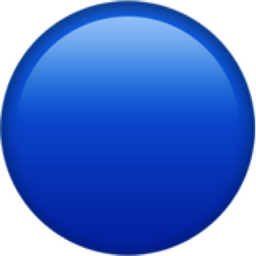}}
  \raisebox{-.125em}{\includegraphics[height=1em]{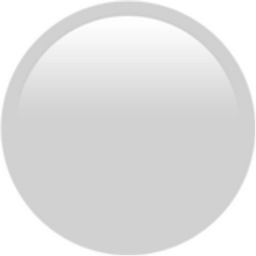}}
  \raisebox{-.125em}{\includegraphics[height=1em]{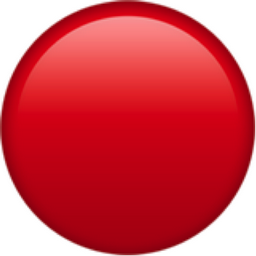}}. Once satisfied with an emoji palette, the author specifies (b) the type and parameters of a unit chart (c) or time series chart (\cref{fig:emoji-timeseries}). The completed chart can appear alongside other charts in the dashboard (3, 4) or be copied and pasted for text-based communication, such as in a Slack Message (5). See the supplemental video to see each chart construction step.}
  \label{fig:teaser}
  \vspace{-0.5em}
}

\abstract{
    \input{sections/00_abstract}
} 


\begin{document}


\firstsection{Introduction}

\maketitle

\input{sections/01_intro}

\input{sections/02_rw}

\input{sections/03_octomoji}


\input{sections/04_discussion}

\section*{Supplemental Materials}
\label{sec:supplemental_materials}

We include a 2.5-minute video demonstration of the \octomoji~extension for Tableau, beginning with the table of data featured in \cref{fig:teaser}.2 and concluding with the resulting pictographic unit chart pasted into a Slack message \cref{fig:teaser}.5.

\acknowledgments{Matthew Brehmer, Zoe, and Michael Correll contributed to this work while they were affiliated with Tableau. We thank Jason Bekkedam for his assistance with the Tableau Dashboard Extensions API, as well as Ana Crisan and Kate Mann for their feedback.}

\bibliographystyle{abbrv-doi}

\bibliography{main}
\end{document}

%% file: sections/00_abstract.tex
Communicating data insights in an accessible and engaging manner to a broader audience remains a significant challenge. To address this problem, we introduce the \octomoji, a tool that generates a set of emoji recommendations for the field and category names appearing in a tabular dataset. The selected set of emoji encodings can be used to generate configurable unit charts that combine plain text and emojis as word-scale graphics. These charts can serve to contrast values across multiple quantitative fields for each row in the data or to communicate trends over time. Any resulting chart is simply a block of text characters, meaning that it can be directly copied into a text message or posted on a communication platform such as Slack or Teams. This work represents a step toward our larger goal of developing novel, fun, and succinct data storytelling experiences that engage those who do not identify as data analysts. Emoji-based unit charts can offer contextual cues related to the data at the center of a conversation on platforms where emoji-rich communication is typical.

%% file: sections/01_intro.tex
Data storytelling can assume many forms~\cite{segel2010narrative}, and some of these forms are more appropriate for certain media than others. 
For instance, magazine-style articles and data comics~\cite{bach2018design} may be better suited for a print medium, while the use of scrollytelling~\cite{mckenna2017visual} and data videos~\cite{amini2015understanding} may be better suited for online journalism and broadcast media, respectively. 
In this paper, we address the medium of \textit{text-based communication}, encompassing text messages, social media posts, blog posts, journalistic articles, and conversations on collaboration platforms such as Slack~\cite{slackStat} or Teams~\cite{teamsStat}.
Distinguishing aspects of this medium include the expressive use of emojis, memes, and animated GIFs to convey a certain tone, effect, or nuance that text alone cannot accomplish succinctly. Our research specifically investigates the potential of emojis for communicative data visualization, particularly due to their universality (there is currently a finite set of emojis in the Unicode standard), their broad familiarity~\cite{emojipedia}, and their expressivity~\cite{khandekar2019opico}.   
We further argue that data storytelling with emojis is appropriate for audiences who do not identify as data analysts and especially those who may be more receptive to fun or playful pictorial representations of data.

Constructing a bespoke emoji-based chart that reflects the underlying semantics of the data is currently possible in any emoji-compatible text editor, as exemplified by the use of emojis in social media posts to communicate proportions~\cite{emojiVB} or timelines~\cite{emojiKM}.
However, this construction process can be tedious. To this end, we apply methods from the artificial intelligence community to generate a palette of semantically-resonant emojis based on the field and category names present in the tabular dataset provided by a chart author. Surfacing these recommended palettes in an interactive authoring interface as part of a Tableau dashboard extension (\cref{fig:teaser}) allows for deeper integration with a chart author's workflow. This \octomoji~interface is the primary contribution of this work. As charts generated using this interface are simply text blocks comprised of alphanumeric and emoji characters, they can easily be copied and pasted for use in any text-based communication, from social media posts and conversation platform threads to inline word-scale graphics~\cite{goffin2014exploring} in documents or articles. 
This functionality allows for succinct and semantically suggestive data storytelling that is appropriate for any text-based medium.

%% file: sections/02_rw.tex
\section{Background \& Related Work}
\label{sec:rw}

Our work builds upon prior work in word-scale information graphics~\cite{goffin2014exploring}, a popular example of which being \textit{sparklines}~\cite{tufte2006beautiful} for representing time-series data. We also take inspiration from pictorial statistics~\cite{burns2021designing} and its rich history, from the use of semantically-suggestive symbols to communicate economic data in Central America prior to European colonization~\cite{navarro}, or the 20th-century development of Isotype~\cite{haroz2015isotype} to draw attention to economic inequality and promote group consciousness amongst the working class~\cite{imagefactories}.

Apart from the manually-constructed emoji charts mentioned above~\cite{emojiKM,emojiVB}, we also draw inspiration from Gamio's emoji-like symbol map of Chernoff Faces~\cite{emojiAxios} and the Dark Sky app's use of emojis in weather reporting~\cite{emojiDarkSky}. 
Both of these instances involve predetermined mappings between quantitative values and visual representations, whereas in our approach, recommended mappings will vary given different fields names and categorical values. 

Prior research in communicative data visualization has employed various NLP techniques to impart some semantic resonance with the underlying data. Techniques include the use of heuristics to populate data marks in charts with thematically-appropriate icons retrieved from the web~\cite{setlur2014automatic} or the use of generative techniques to produce thematically-suggestive data glyphs~\cite{ying2022metaglyph} and mark textures~\cite{wu2023viz2viz,xiao2023let}.
We continue this line of research by applying a neural network in the construction of emoji-based unit charts. 

%% file: sections/03_octomoji.tex
\section{\octomoji}
\label{sec:octomoji}

\octomoji~leverages Word2Vec~\cite{mikolov2013distributed} to generate a set of emoji recommendations for the field and category names appearing in a tabular dataset. We trained the Word2Vec model on emoji descriptions from Emojipedia~\cite{emojipedia}, which has curated paragraph-length descriptions for many of the 3,782 emojis currently included in the Unicode standard (v 15.1). 

\begin{figure}[h!]
 \centering 
 \includegraphics[width=\columnwidth]{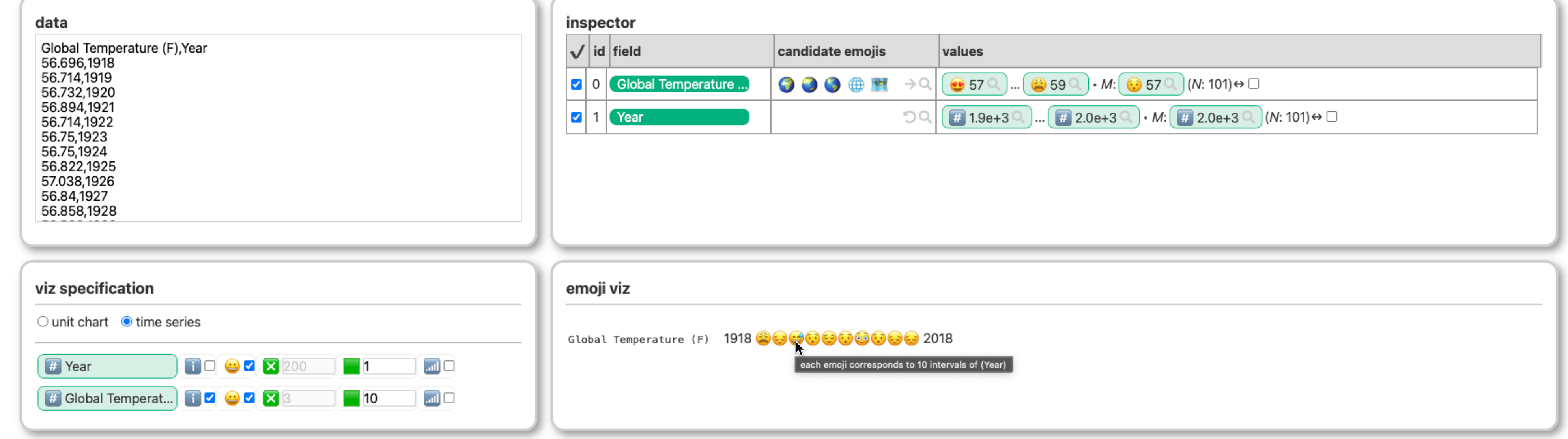}
 \caption{In contrast to the unit chart specification (\cf \cref{fig:teaser}.1.b), the \octomoji~can also be used to represent univariate time series, such as in this instance depicting global average temperature differences over ten-year intervals between 1918 and 2018 using our \textsl{emoji-10} scale: 
 \raisebox{-.125em}{\includegraphics[height=1em]{figures/emoji1.png}}
 \raisebox{-.125em}{\includegraphics[height=1em]{figures/emoji3.png}}
 \raisebox{-.125em}{\includegraphics[height=1em]{figures/emoji6.png}}
 \raisebox{-.125em}{\includegraphics[height=1em]{figures/emoji5.png}}
 \raisebox{-.125em}{\includegraphics[height=1em]{figures/emoji8.png}}
 \raisebox{-.125em}{\includegraphics[height=1em]{figures/emoji5.png}}
 \raisebox{-.125em}{\includegraphics[height=1em]{figures/emoji4.png}}
 \raisebox{-.125em}{\includegraphics[height=1em]{figures/emoji5.png}}
 \raisebox{-.125em}{\includegraphics[height=1em]{figures/emoji3.png}}
 \raisebox{-.125em}{\includegraphics[height=1em]{figures/emoji3.png}}.
 }
 \label{fig:emoji-timeseries}
 \vspace{-0.5em}
\end{figure}

A chart author can provide tabular data to the \octomoji~ in two ways, either by ingesting data from a CSV table, such as the annual average global temperature differential values shown in \cref{fig:emoji-timeseries}, or by referencing tabular data in a Tableau dashboard, such as the table shown in \cref{fig:teaser}.2: flood risk values across the Netherlands, along with the number and type of employees in each province. 
Once ingested, the \octomoji~authoring interface surfaces emoji recommendations in the inspector interface (\cref{fig:teaser}.1.a), which includes a paginated ranked list of candidate emojis for each field, as well as the top recommended emoji for categorical field values.
If the model does not provide a recommendation, the field or value is assigned a categorical placeholder emoji, and at any point in the authoring process, a recommended or placeholder emoji assignment can be manually overridden by the author via an emoji search interface.
In \cref{fig:teaser}.1.a, \raisebox{-.125em}{\includegraphics[height=1em]{figures/city.png}} and \raisebox{-.125em}{\includegraphics[height=1em]{figures/wave.png}} are among the recommendations for the \field{\texttt{\% cities / towns at risk of rising water levels}} field, while \raisebox{-.125em}{\includegraphics[height=1em]{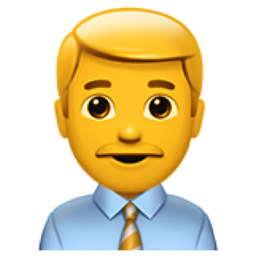}} is among the top recommendations for the \field{\texttt{number of remote workers}} field.
For numerical field values, we provide a menu of ordinal emoji palettes, such as our `\textit{emoji-10}' scale
(\raisebox{-.125em}{\includegraphics[height=1em]{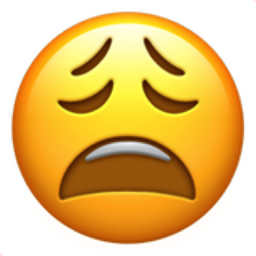}}
\raisebox{-.125em}{\includegraphics[height=1em]{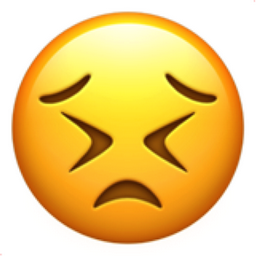}}
\raisebox{-.125em}{\includegraphics[height=1em]{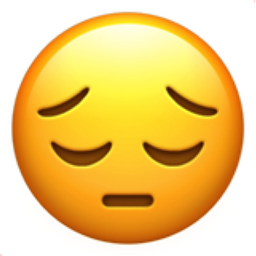}}
\raisebox{-.125em}{\includegraphics[height=1em]{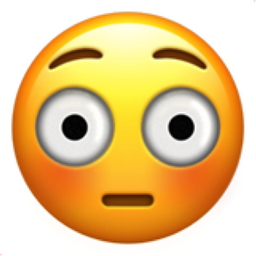}}
\raisebox{-.125em}{\includegraphics[height=1em]{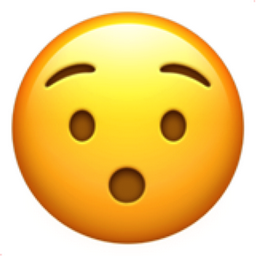}}
\raisebox{-.125em}{\includegraphics[height=1em]{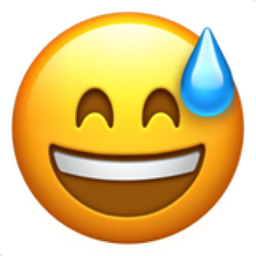}}
\raisebox{-.125em}{\includegraphics[height=1em]{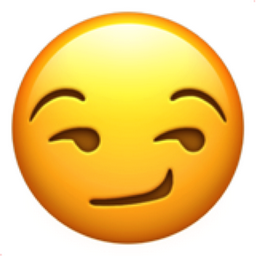}}
\raisebox{-.125em}{\includegraphics[height=1em]{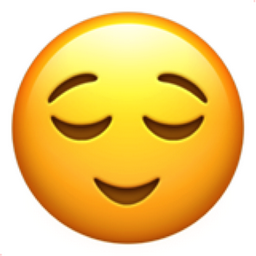}}
\raisebox{-.125em}{\includegraphics[height=1em]{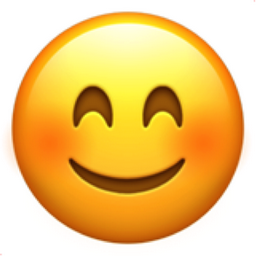}}
\raisebox{-.125em}{\includegraphics[height=1em]{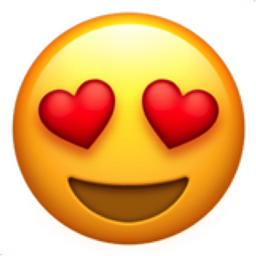}}) 
used in \cref{fig:emoji-timeseries} or the diverging 
\raisebox{-.125em}{\includegraphics[height=1em]{figures/blue.png}}
\raisebox{-.125em}{\includegraphics[height=1em]{figures/white.png}}
\raisebox{-.125em}{\includegraphics[height=1em]{figures/red.png}} 
scale used in \cref{fig:teaser}.1.a.i. 

The selected set of emoji assignments for field names and values are automatically propagated to a chart specification panel (\cref{fig:teaser}.1.b).
Choices that an author makes using this panel are in turn reflected in a preview of the emoji chart in the lower right panel.  
The chart specification panel includes template configurations for stacked bar unit charts (\cref{fig:teaser}.1.c) and univariate time series charts (\cref{fig:emoji-timeseries}).
The former configuration includes affordances for specifying aggregation and grouping factors for numerical fields, while the latter allows for specifying a window size for the provided temporal field. 
Lastly, the author re-arrange the ordering of fields via drag and drop and opt to include field and value names as plain text.
Once satisfied, the authoring interface can be dismissed, and the resulting chart can appear alongside other dashboard elements (\cref{fig:teaser}.3 and \cref{fig:teaser}.4) or be copied to the author's clipboard as text and pasted into any text-based communication, such as a Slack message (\cref{fig:teaser}.5), social media post, or document.

%% file: sections/04_discussion.tex
\section{Discussion \& Future Work}
\label{sec:discussion}
Many chart construction interfaces assume that those using them already have a sense of how they want to represent their data~\cite{satyanarayan2019critical}; in other words, they are authoring tools rather than design tools. 
The \octomoji, like other recent interfaces that incorporate AI techniques to generate semantically-resonant visual encodings (\eg~\cite{wu2023viz2viz,xiao2023let,ying2022metaglyph}), represents a step toward closing the gap between visualization design and authoring.

Immediate next steps include conducting human-centered evaluations. First, prior reflections on the evaluation of chart authoring interfaces~\cite{Amini2018DDS,ren2018reflecting} suggest that a chart reproduction or replication study could shed light on the interface's learnability and usability.
Second, the results of an open-ended chart construction activity with participants' own data could provide us with a sense of the potential utility of our approach, particularly if complemented by retrospective interviews with participants as well as with the audiences for their charts.
Finally, we can look to methodological precedents for evaluating the readability of pictographic representations~\cite{burns2021designing,haroz2015isotype} to conduct similar experiments that specifically focus on the emoji-based charts constructed using the \octomoji. 

Another direction for future work is to consider additional recent advances in AI for generating a broader variety of emoji-rich representations: 
(\textbf{1}) Beyond unit charts for comparing quantities and sparkline-inspired emoji sequences for time series, symbol maps and set diagrams are two particularly alluring possibilities. 
(\textbf{2}) We could evaluate the utility of large language models to generate explanatory narratives around patterns observed in the data: text-based explanations that embed recommended or selected emojis throughout as word-scale graphics~\cite{goffin2014exploring}.
(\textbf{3}) We might use these models to generate additional ordinal emoji palettes. 
(\textbf{4}) We could replace or complement the current template-based chart specification interface (\cref{fig:teaser}.1.b) with a prompt-based interface for describing and adjusting the layout of an emoji-based chart, or a multimodal interface that replaces abstract visual elements in an existing chart image with emojis. 
(\textbf{5}) Lastly, if we relax the constraints of using existing emojis and generating entirely text-based output, we could generate palettes of novel or hybrid emoji-based icons resulting from traversals of the latent space between emojis~\cite{liu2019latent}.

Irrespective of whether recent generative models or prior approaches are used to construct emoji-based representations of data, the use of emojis has two 
additional benefits that merit discussion. 
First, relative to chart construction techniques that leverage text-to-image models, emoji-based charts are more likely to be stylistically consistent without significant refinement. 
Second, charts constructed using text-to-image models could  (intentionally or unintentionally) incorporate styles, motifs, or visual elements associated with particular artists who did not consent to their works being used in this way; in contrast, all Unicode emojis are in the public domain, alleviating some potential ethical concerns when publishing or sharing an emoji-based chart.